\documentclass[useAMS,usenatbib]{mn2e}
\usepackage{graphicx}
\usepackage[euler]{textgreek}
\usepackage{breqn}
\usepackage{amssymb}
\usepackage{color}
\usepackage{comment}
\usepackage{subfigure}

\title[HE neutrino emission from the core of LLAGNs triggered by magnetic reconnection]{High energy neutrino emission from the core of low luminosity AGNs triggered by magnetic reconnection acceleration}
\author[B. Khiali, E. M. de Gouveia Dal Pino]{B. Khiali$^{1}$\thanks{E-mail:
bkhiali@usp.br} and E. M. de Gouveia Dal Pino$^{1}$\\
$^{1}$IAG-Universidade de S\~ao Paulo, Rua do Mat\~ao 1226, S\~ao Paulo, SP, Brazil}

\begin{document}

\date{Accepted 2015 October 6. Received 2015 September 21; in original form 2015 June 02}

\pagerange{\pageref{firstpage}--\pageref{lastpage}} \pubyear{2015}

\maketitle

\label{firstpage}

\begin{abstract}

The detection of astrophysical very high energy (VHE) neutrinos in the  range of TeV-PeV energies by the IceCube observatory has opened a new season in high energy astrophysics. Energies $\sim$PeV imply that the neutrinos are originated from sources where cosmic rays (CRs) can be accelerated up to $\sim 10^{17}$eV. Recently, we have shown that the observed TeV gamma-rays from radio-galaxies may have a hadronic origin in their nuclear region and in such a case this could lead to neutrino production. In this paper we show that relativistic protons accelerated by magnetic reconnection in the core region of these sources may produce HE neutrinos via the decay of charged pions produced by photo-meson process. We have also calculated the diffuse flux of HE neutrinos and found that it can be associated to the IceCube data.

\end{abstract}


\section{Introduction}

Neutrino observations can provide unique information to understand their origin and can even lead to the discovery  of new classes of astrophysical sources. 
The inherent isotropic nature of the detected neutrino flux by IceCube is compatible with an  extragalactic origin and is supported by diffuse high energy $\gamma$-ray data (\citealt{ahlers14}). The observed  neutrinos with energies $\sim PeV$ suggest that they are originated from a source where cosmic rays (CRs) can be accelerated up to $\sim10^{17} \rm{eV}$.

A potential  mechanism to produce  VHE neutrinos  in the TeV-PeV range is through the decay of charged pions created in proton-proton  ($pp$) or proton-photon ($p\gamma$) collisions in a variety of astrophysical sources which, in the framework of the IceCube observations, may include  active galactic nuclei (AGNs) (\citealt{kasanas86,stecker91,atoyan01,neronov02}) and gamma-ray bursts (GRBs) (\citealt{waxman97}).

Hadronic mechanisms producing VHE neutrinos via the acceleration of cosmic rays (CRs)  in AGNs   have been  suggested for more than three decades (\citealt{eichler79, protheroe83, mannheim95, hazlen97, mucke01, kalashev14,marinelli14b,atoyandermer2003,becker08}). Currently, the detection of gamma-ray emission at TeV energies in AGNs, not only in  high luminous blazars,  but also in less luminous radio-galaxies,  has strengthened the notion that they may be excellent cosmic ray accelerators and therefore, important potential  neutrino emission candidates.

Several recent models have tried to describe the detected TeV neutrino emission as due to AGNs. 
For instance, \cite{marinelli14b} employed  two different hadronic scenarios involving the interaction of accelerated protons at the AGN jet either with photons produced via synchrotron self-Compton (SSC)  or with thermal particles in the giant lobes.  They then derived  the expected neutrino flux  for  low luminous AGNs (LLAGNs) \footnote{By LLAGNs we mean non-blazar sources with $L_{H_\alpha}\leq10^{40} \rm{erg s^{-1}}$ (see \citealt{ho97,nagar05}), where $L_{H_\alpha}$ is the $H_\alpha$ luminosity. These  typically consist of liners and seyfert galaxies which are also  FR I or FR II radio sources. For more details see \cite{beteluis2014}.}, or more specifically, for radio galaxies for which they examined the origin of the observed  TeV gamma-ray spectra as due to  hadronic processes.
 
Earlier work by \cite{gupta08}  had already introduced  hadronic scenarios to explain the TeV emission in LLAGNs (e.g., Cen A). Also, \cite{fraija14a,fraija14b} suggested  neutral pion decays from $pp$ and $p\gamma$ interactions in these sources as  probable candidates to explain the high energy neutrinos. 
In another model, \cite{kalashev14} attempt to reproduce the IceCube data using the $p\gamma$ mechanism considering the radiation field produced by the accretion disk around the   AGN central black hole (assuming a standard Shakura-Sunyaev accretion disk model). 
Alternatively, \cite{kimura14} calculated the neutrino spectra using the radiatively inefficient accretion flows (RIAF) model in the nuclei of LLAGNs considering both $pp$ and $p\gamma$ mechanisms and  stochastic proton acceleration in the RIAF turbulence.
 
The possibility of producing VHE neutrino emission  has been also extensively explored in blazars - AGNs for which the relativistic jet points to the line of sight (e.g., \citealt{atoyandermer2003,becker08,murase14,dermer14}).  \cite{dermer14}, in particular, revisited the previous  studies assuming that the observed neutrinos could be produced in the inner jet of blazars and concluded that neither the flux nor the spectral shape suggested by the IceCube data could be reproduced by this scenario which predicts a rapid decline of the emission below 1 PeV. 
\citealt{tavecchio14} and \citealt{tavecchio14b}, on the other hand, considered the distribution of lower-power blazars, namely, BL Lac objects and, by employing a two-zone spine-sheath jet model to these sources concluded  that they might be  suitable for the production of the observed PeV neutrinos  revealed by the IceCube.

Presently, it is very hard to define what should be the dominant process or the real sources that are producing  the observed neutrino flux mainly  due to the lack of more precise measurements. But while  waiting for better measurements, we can explore further  mechanisms and try to make reliable predictions in order to constrain the candidates. 

The big challenge in  models that rely  on hadronic processes in the AGN nuclei is how to produce the relativistic protons that may lead to $\gamma$-ray emission and the accompanying neutrino flux. 
Diffusive shock acceleration  at the jet launching base was discussed by \cite{begelman90}. \cite{levinson} and more recently, \cite{vincent} proposed that  
TeV gamma-ray emission might be produced in the BH magnetosphere by  pulsar-like  mechanisms, i.e., with particles being accelerated by the electric potential difference settled by non uniform magnetic field.
 As remarked above, \cite{kimura14} discussed stochastic acceleration in an accreting RIAF turbulent scenario, but currently none of these models can be regarded as dominant  or disclaimed given the uncertainties from  the observations regarding the location of the gamma-ray emission (see \S.~5).

In this work, we consider an alternative acceleration model that may occur in the vicinity of BHs which was  explored first in the framework of microquasars by \citealt{gl05} (hereafter GL05) and then  extended  to  AGNs by \citealt{beteluis2010a} (hereafter GPK10). In this model, particles can be accelerated in the surrounds of the BH by the magnetic  power extracted from events of fast magnetic reconnection occurring between the magnetosphere of the BH  and the lines rising from the inner accretion disk into the corona (Figure 1).

 More recently,  Kadowaki, de Gouveia Dal Pino \& Singh (2015)  revisited this model exploring different mechanisms of fast magnetic reconnection and extending the study to include also the gamma-ray emission of a large sample of sources (more than 200 sources involving blazars, non-blazars or LLAGNs, and galactic black hole binaries). They confirmed the earlier trend found by GL05 and GPK10, verifying that  the fast magnetic reconnection power calculated as a function of the black hole (BH) mass can explain the observed radio and gamma-ray luminosity from nuclear outbursts of all LLAGNs and galactic black hole binaries of their sample, spanning $10^{10}$ orders of magnitude in mass  (see  Fig. 5 in \citealt{beteluis2014}).
 \footnote{We note that the calculated reconnection power in this core model though  large enough to explain the luminosity of galactic black hole binaries and LLAGNs (or non-blazars), is insufficient for reproducing the luminosity of most of the blazars and  GRBs, which is compatible with the notion that the observed emission in these cases is produced outside the core, at the jet that points to the line of sight and screens any deep nuclear emission (\citealt{beteluis2014}).}

In the works  above, a standard accretion disk model was employed, but in an accompanying work (\citealt{chandra14}), the authors adopted a  magnetically dominated advective flow (MDAF) for the accretion and obtained very similar results to those above, which demonstrated that  the
details of the accretion physics are not relevant in the development of the magnetic reconnection process and the particle acceleration occurring in the corona.

The magnetic reconnection acceleration  model above (see \citealt{beteluis2014})   has been also employed to reproduce the observed  multi-wavelength spectral energy distribution (SED) of  a few microquasars (Cygnus X-1 and Cygnus X-3)  (Khiali, de Gouveia Dal Pino \& del Valle (2015), henceforth KGV15) and LLAGNs (Cen A, NGC 1275, M87 and IC310)  (Khiali, de Gouveia Dal Pino \& Sol (2015), hereafter KGS15), and the results indicate that hadronic mechanisms ($pp$ and $p\gamma$) are  the main radiative processes producing the observed GeV to TeV $\gamma$-rays.
 
The results of the works above  and in particular, the reproduction of the observed  SEDs and the TeV  gamma-ray) emission of the radio-galaxies by hadronic processes involving  particles accelerated by magnetic reconnection  in the surrounds of the BH (KGS15), have  motivated  the present study. 
We aim here to calculate the spectrum of neutrinos arising from the interactions of 
 accelerated protons by the mechanism above with  the radiation and thermal-particle  fields around the BH. According to our previous results (KGV15 and KGS15),  these interactions produce weakly decaying $\pi^0$ and  $\pi^{\pm}$ pions.  The latter may generate high energy neutrinos.
  We will then evaluate
 the diffuse neutrino intensity and compare with the IceCube data in the context of LLAGNs.
  
For completeness, we will  also compare the particle acceleration by magnetic reconnection with shock acceleration in the surrounds of the BH.  

The outline of the paper is as follows. In Section 2, we describe in detail our scenario. In Section 3, we describe the hadronic interactions and calculate the acceleration and radiative cooling rates. The calculation of the spectrum of neutrinos and their diffuse intensity for  comparison with the IceCube data is presented in Section 4. We discuss and summarize our results and conclusions in Section 5.

\section{Description of the model}

In this section we summarize the  main characteristics of our fast magnetic reconnection model in the surrounds of the BH and  how particles can be accelerated in the magnetic reconnection layer. For a more detailed description we refer to \citealt{beteluis2014}.

\subsection{Fast magnetic reconnection in the surrounds of the BH}

\begin{figure}
 \centering

 \includegraphics[width=3in]{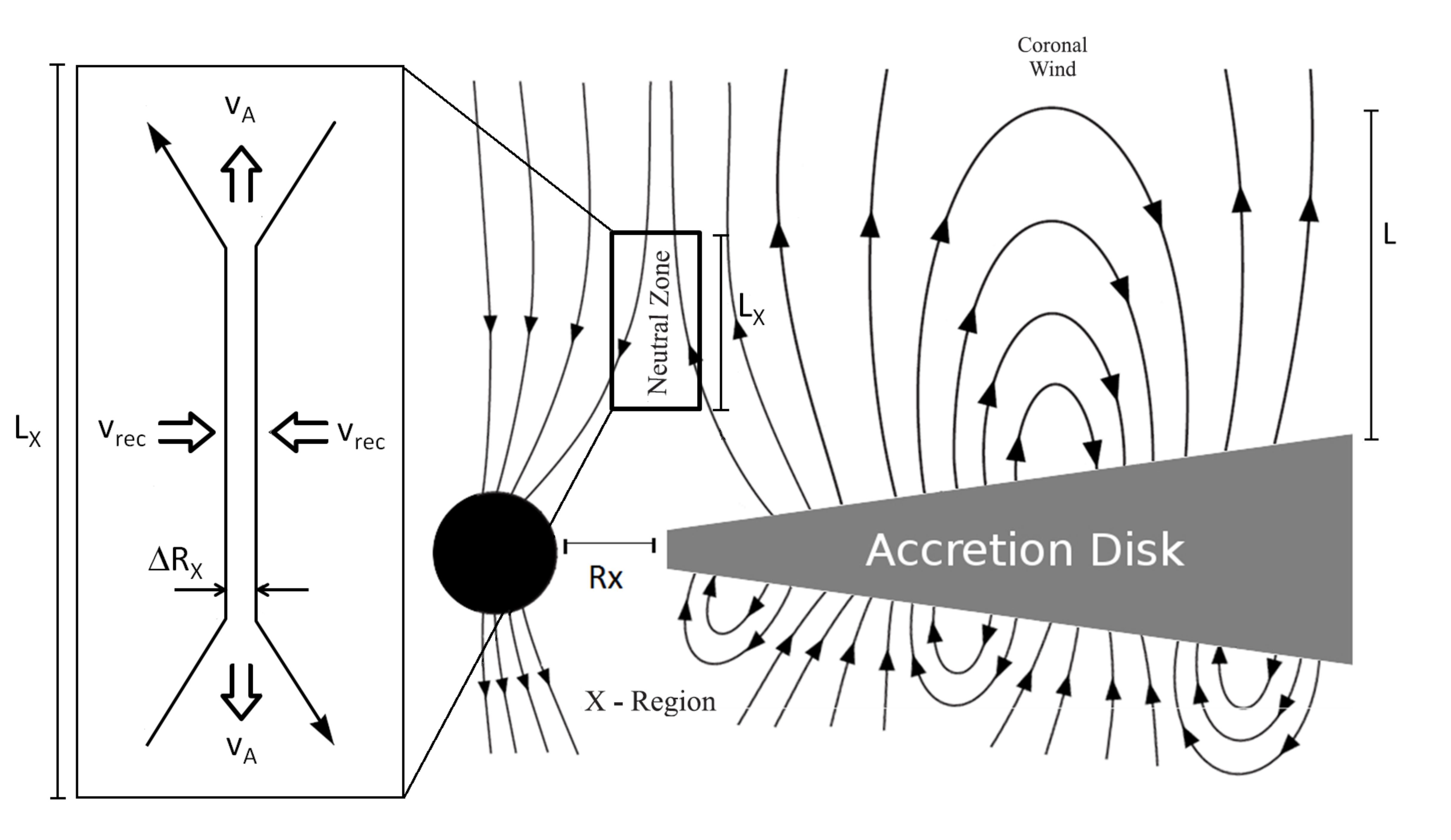}
 \caption{Scheme of magnetic reconnection between the lines arising from the accretion disk and the lines anchored into the BH horizon. Reconnection is made fast by the presence of embedded turbulence in the reconnection (neutral) zone (see text for more details).  Particle acceleration may occur in the magnetic reconnection zone by a first-order Fermi process (adapted from GL05). }
 \label{fig1}
\end{figure}

We assume that the gamma-ray emission from low-luminous AGNs is produced in  the core region and the particles responsible for this emission are accelerated in the corona around the BH and accretion disk as sketched in Figure~\ref{fig1}.

Turbulent dynamo inside the accretion disk or plasma dragging from the surrounds can  build the large-scale poloidal magnetic fields that arise into the corona. This poloidal magnetic flux under  the action of  disk differential rotation gives rise to a wind that partially removes angular momentum from the system  increasing the accretion rate and  the ram pressure of the accreting material that will then press the magnetic lines in the inner disk region against the lines anchored into the BH horizon allowing them to reconnect (see Fig. 1). The magnetic field intensity in the inner region of the accretion disk corona is  approximately given  by the balance   between the magnetic pressure of the BH magnetosphere and the accretion ram pressure (\citealt{beteluis2014}):
\begin{equation} \label{1}
B\cong 9.96\times 10^{8}r_X^{-1.25} \xi^{0.5}m^{-0.5}\ {\rm G}.
\end{equation}
Where $r_X=R_X/R_S$ is the inner radius of the accretion disk in units of the BH Schwartzchild radius ($R_S$) (in our calculations we assume $r_X=6$); $\xi$ is the  mass accretion disk rate in units of the Eddington rate ($\xi=\dot{M}/\dot{M}_{Edd}$), with  $\dot{M}_{Edd} =1.45\times10^{18}  m$ g s$^{-1}$), which  we  assume to be  $\xi \simeq 0.7$\footnote{See Fig. 5 in \cite{beteluis2014}. Accretion rates $\xi$  between $0.05 < \xi \leq  1$ are able to produce magnetic reconnection power values which are enough to probe the observed gamma-ray luminosities from LLAGNs.}; $m$ is the BH mass in units of solar mass.

The presence of embedded turbulence in the nearly collisional MHD coronal flow of the core region of the AGNs can make reconnection   very  fast with a rate $V_R \simeq v_A (l_{inj} /L)^{1/2} (v_{turb}/v_A)^2$, where  $l_{inj}$ and $v_{turb}$ are the injection scale and velocity of the turbulence, respectively (e.g., \citealt{LV99}, hereafter LV99\footnote{According to the LV99 model, even weak embedded turbulence causes the wandering of the magnetic field lines which allows for many independent patches to reconnect simultaneously making the global reconnection rate  large (for more details see \citealt{beteluis2014}). This theory has been confirmed numerically by means of 3D MHD simulations (\citealt{kowal09,kowal2012}).}.
 This relation shows  that the reconnection rate is of the order of the Alfv\'en speed $v_A$, which
in the systems here considered may be near the light speed.

The magnetic reconnection power released by turbulent
driven fast reconnection in the magnetic discontinuity region
(Figure~\ref{fig1})  is given by  (\citealt{beteluis2014}):
\begin{equation} \label{2}
W \simeq 1.66\times 10^{35} \psi^{-0.5} r_X^{-0.62} l^{-0.25} l_X q^{-2}\xi^{0.75}m\ ~ {\rm erg~s^{-1}},
\end{equation}
where  $l=L/{R_S}$ is the height of the corona in units of $R_S$; $l_X={L_X}/{R_S}$, $L_X$ is the extension of the magnetic reconnection zone (as shown in Figure~\ref{fig1}), $q=[1-(3/r_X)^{0.5}]^{0.25}$ and $v_A = v_{A0} \psi$ is the relativistic form of the Alfv\'en velocity,   with $v_{A0} = B/(4\pi \rho)^{1/2}$,  $B$ being the local magnetic field, $\rho \simeq n_c m_p$  the fluid density, and  $\psi=[1+({v_{A0} \over c})^2]^{-1/2}$, in this  work,  we find that  $v_{A0} \sim c$.

The acceleration region in our model corresponds to the cylindrical shell around the BH where magnetic reconnection takes place, as shown in Figure~\ref{fig1}. This shell has a length $l_X$, with  inner and outer radii  given by $R_X$ and $R_X+\Delta R_X$ respectively, where $\Delta R_X$ is the width of the current sheet  given by (\citealt{beteluis2014}): 
\begin{equation} \label{3}
\Delta R_X\cong 2.34\times 10^{4} \psi^{-0.31} r_X^{0.48} l^{-0.15} l_X q^{-0.75}\xi^{-0.15}m\ {\rm cm}.
\end{equation}

This magnetic reconnection  power (Eq.~\ref{2}) will  both heat the surrounding gas and accelerate particles. As in \cite{beteluis2014}, we assume that approximately 50\% of the reconnection power goes to accelerate the particles (see \S. 2.2.).  This is consistent with plasma laboratory experiments of reconnection acceleration (\citealt{yamada}) and also with the observations of solar flares (e.g., \citealt{lin71}). We further assume that this  power is equally shared between the protons and electrons/positrons, so that the proton luminosity will be 25\% of the calculated value by Eq.~\ref{2}.

The particle density  in the coronal region in the surrounds of the BH is (\citealt{beteluis2014}): 
\begin{equation} \label{nc}
n_c\cong 8.02\times 10^{18}r_X^{-0.375} \psi^{0.5} l^{-0.75}q^{-2}\xi^{0.25}m^{-1}\ {\rm cm^{-3}}.
\end{equation}

\subsection{Particle acceleration due to the magnetic power released by fast reconnection}
 
In the magnetic reconnection layer (or current sheet; see Figure 1)  where the two converging magnetic flux tubes move to each other with a velocity $V_{R}$, trapped particles may bounce back and forth  due to head-on collisions with magnetic fluctuations in the current sheet. As a consequence,  their energy after a round trip may increase by $< \Delta E/E > \sim  V_{R}/c$, implying an exponential energy increase after several round trips.  This   first-order Fermi acceleration process within reconnection layers was first studied by GL05 and successfully 
 tested through 3D MHD simulations with test particles injected in current sheets with fast reconnection driven by turbulence (\citealt{kowal2011,kowal2012}; see also (\citealt{bete14,betegreg15} for  reviews)\footnote{Particle acceleration within reconnection sheets has been also extensively tested numerically  in collisionless fluids by means of 2D (e.g.,   \citealt{zenitani01,drake06,drake10,cerutti13}) and 3D particle in cell simulations (\citealt{sironi14}).}.

From the results of the 3D MHD numerical simulations (Kowal et al. 2012), we find  that the acceleration rate for a proton is given by (see also KGV15):
\begin{equation}\label{5}
t^{-1}_{acc,M.R.,p}=1.3\times 10^5\left(\frac{E}{E_0}\right)^{-0.4}t_0^{-1},
\end{equation}
where $E$ is the energy of the accelerated proton, $E_0=m_p c^2$, $m_p$ is the proton rest mass, $t_0=l_{acc}/{v_A}$  is the Alfv\'en time,  and $l_{acc}$ is the length scale of the acceleration region and for electrons  this rate is (KGV15):
\begin{equation}\label{6}
t^{-1}_{acc,M.R.,e}=1.3\times 10^5\sqrt{\frac{m_p}{m_e}}\left(\frac{E}{E_0}\right)^{-0.4}t_0^{-1},
\end{equation}
where $m_e$ is the electron rest mass.

As stressed in GL05 (see also KGV15 and KGS15), it is also possible that a diffusive shock may develop in the surrounds of the magnetic reconnection zone due to coronal mass ejections released by fast reconnection along the  magnetic field lines, as observed in solar flares. A similar picture has been also suggested by e.g., \cite{romerovieyro2010}. In this case, the shock velocity  will be predominantly parallel to the magnetic field lines and the  acceleration rate  for a particle of energy $E$ in a magnetic field $B$, will be approximately given by  (e.g.,  \citealt{spruit88}):
 \begin{equation}\label{7}
 t^{-1}_{acc,shock}=\frac{\eta e c B}{E},
 \end{equation}
where $0<\eta \ll 1$ characterizes the efficiency of the acceleration. We fix $\eta=10^{-2}$, which is appropriate for shocks with velocity $v_s\approx 0.1c$, which are  commonly assumed in the Bohm regime  (\citealt{romerovieyro2010}).

\section{Hadronic interactions}

\begin{figure*} 
    \centering
    \subfigure[$Model~1$]
    {
        \includegraphics[width=3.2in]{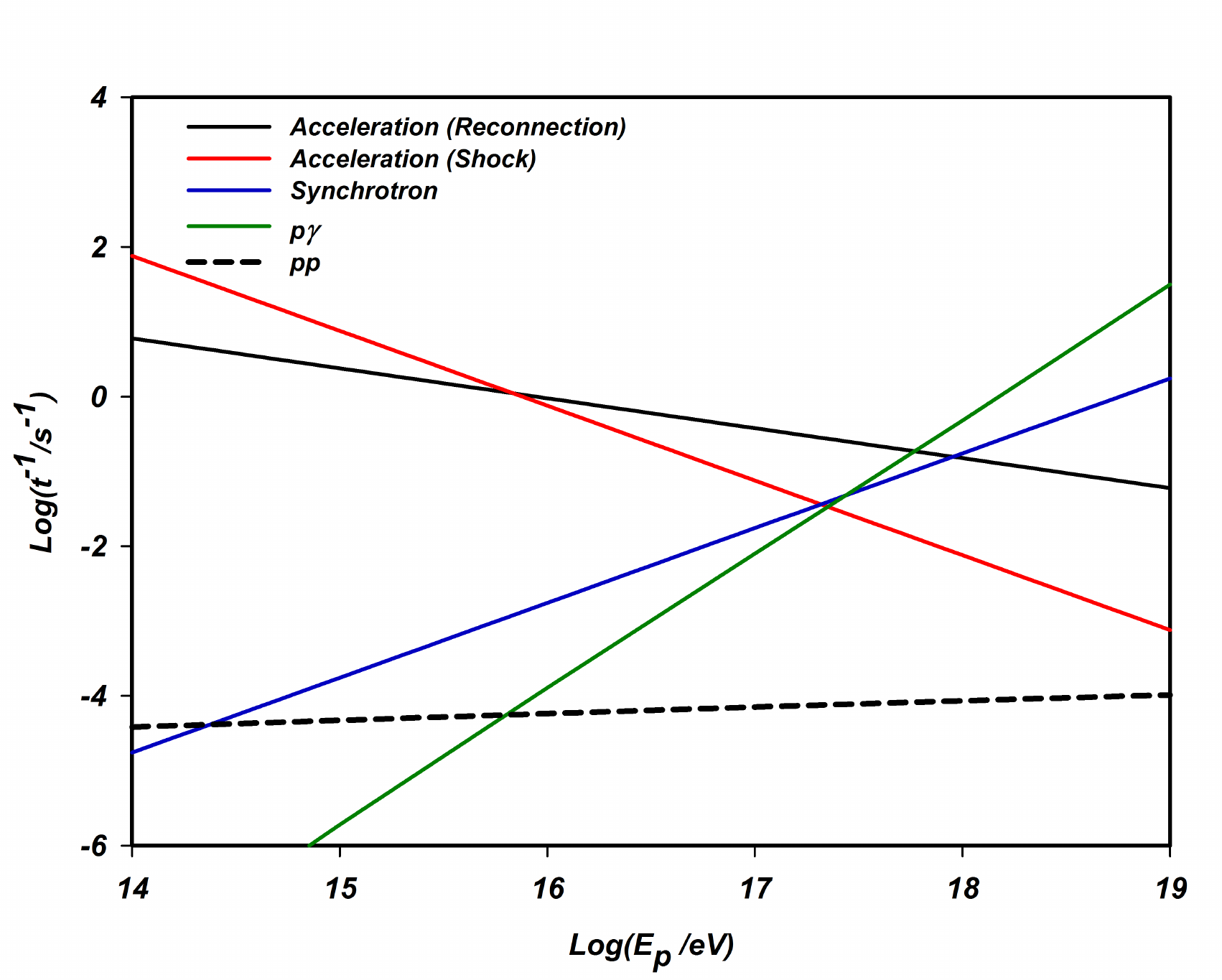}
        \label{2a}
    }
    \subfigure[$Model~2$]
    {
        \includegraphics[width=3.2in]{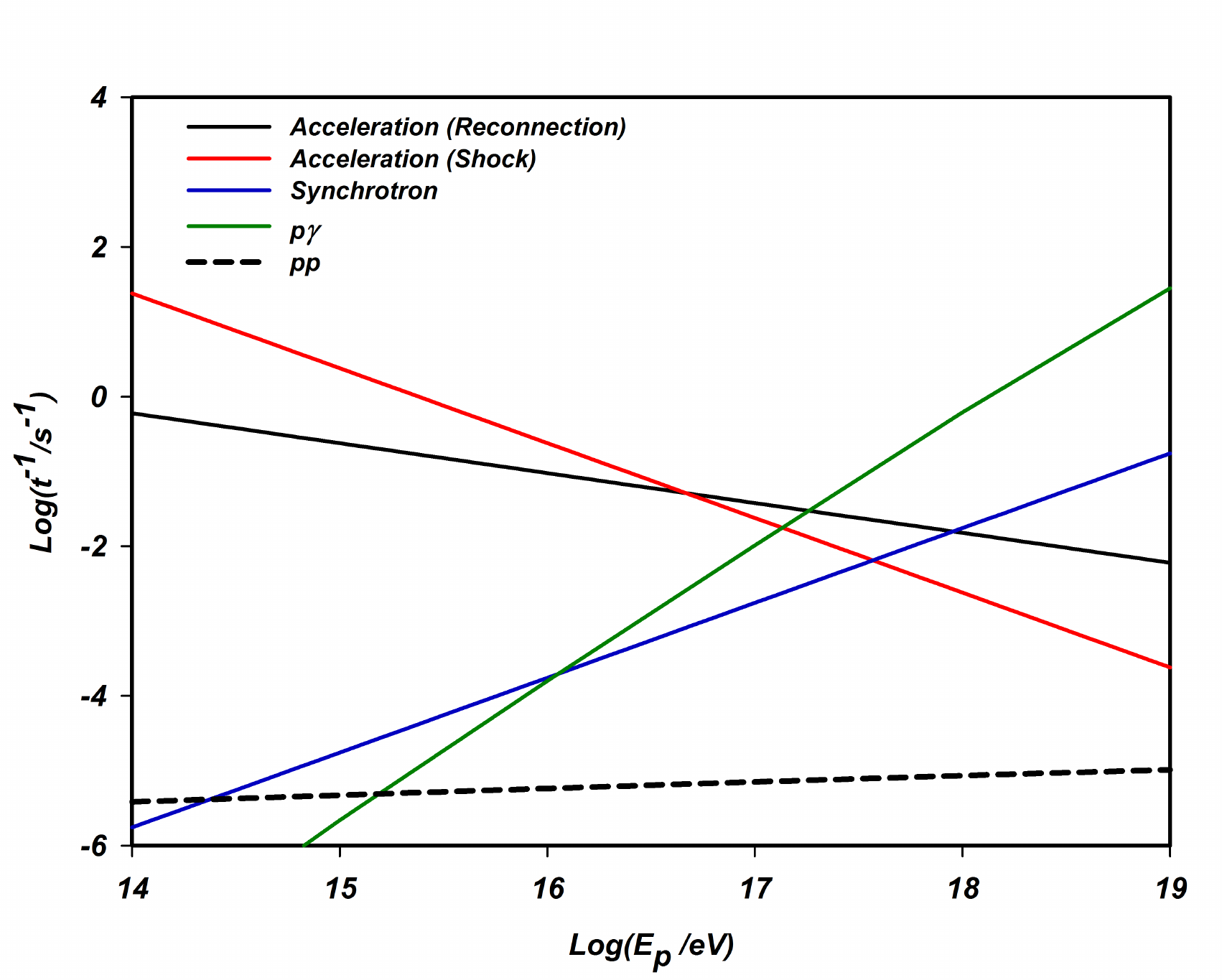}
        \label{2c}
    }
    \\
    \subfigure[$Model~3$]
    {
        \includegraphics[width=3.2in]{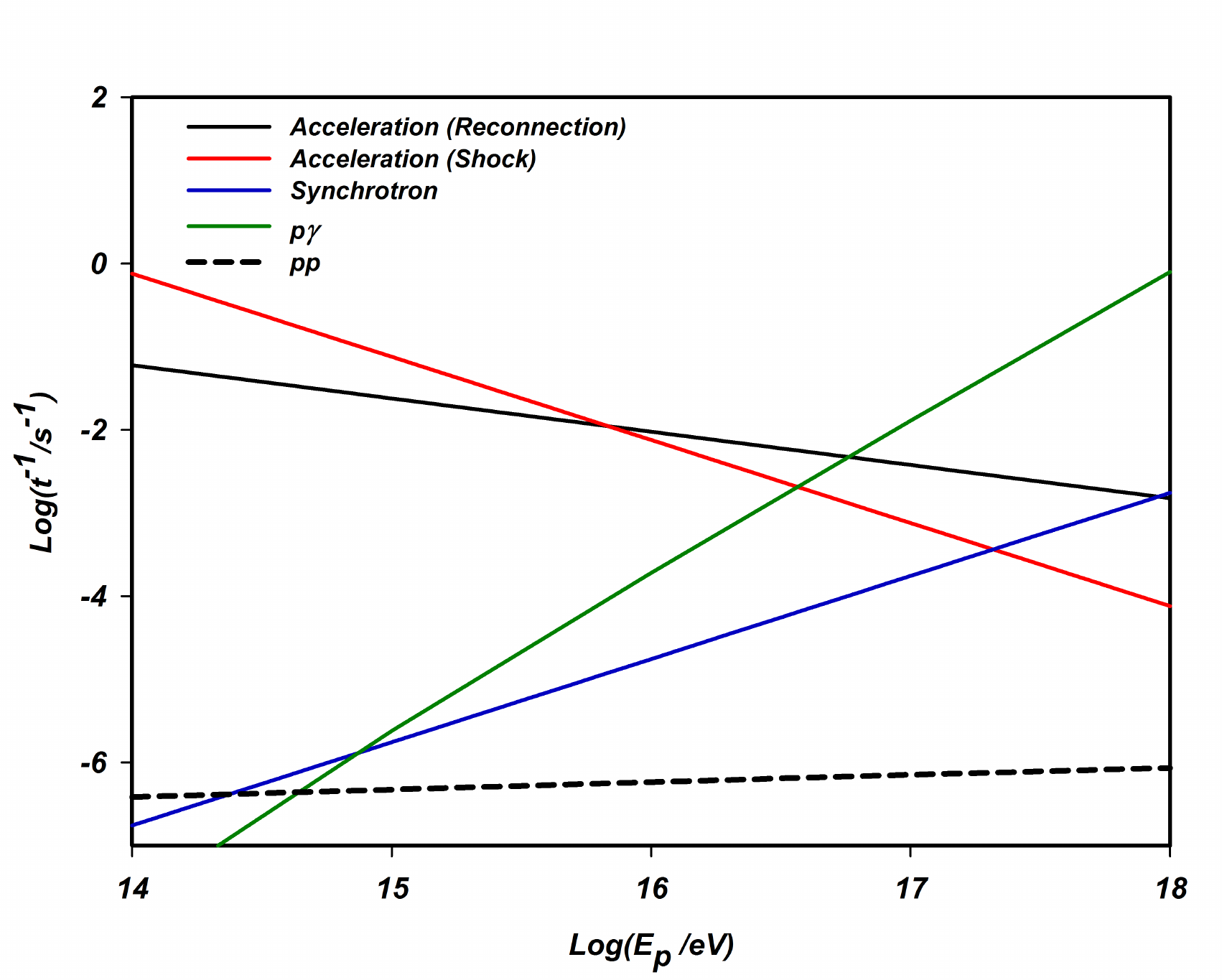}
        \label{2e}
    }
    \caption{Acceleration and cooling rates for  protons in the core regions of LLAGNs with a central black hole mass ($a$)  $M=10^7M_{\odot}$ (Model 1), ($b$)  $M=10^8M_{\odot}$ (Model 2), and ($c$)  $M=10^9M_{\odot}$ (Model 3).
    }
    \label{fig2}
\end{figure*}

In KGS15, we have demonstrated that the core region of LLAGNs is able to accelerate protons  up to energies of a few $10^{17}$eV  through the first-order Fermi magnetic reconnection mechanism described in the previous section. This indicates that these sources could be powerful CR accelerators. 
We show below that these protons can cool very efficiently via synchrotron, $p\gamma$ and $pp$ interactions  in the region  that surrounds the BH of these sources  (Figure~\ref{fig2}).
As remarked, these hadronic interactions lead to the production of HE $\gamma$-rays and HE neutrinos via decays of neutral and charged pions, respectively. In KGS15,  we have calculated the spectral energy distribution of the HE $\gamma$-ray emission for the  LLAGNs for which this emission has been detected. Below,  we calculate the HE neutrino emission from the nuclear region of these sources.

\subsection{$pp$ collisions}  
The charged pions can be created through inelastic collisions of the relativistic protons with nuclei of the corona that surrounds the BH and the accretion disk by means of the following  reactions (\citealt{atoyandermer2003,becker08})  

\begin{equation}\label{9}
p+p\rightarrow n_1(\pi^++\pi^-)+n_2\pi^0+p+p 
\end{equation}
where $n_1$ and $n_2$ are multiplicities, $\pi^0\rightarrow \gamma+\gamma$ (\citealt{stecker70,stecker71}), carrying $33\%$ of the accelerated proton's energy. The charged pions  $\pi^{\pm}$ then decay and produce neutrinos via $\pi^+\rightarrow \nu_{\mu}+\bar{\nu}_{\mu}+\nu_e+e^+$ and $\pi^-\rightarrow \nu_{\mu}+\bar{\nu}_{\mu}+\bar{\nu}_e+e^-$, where $\nu_{\mu}$, $\bar{\nu}_{\mu}$, and $\nu_e$ are the muon neutrino, muon antineutrino, and electron neutrino, respectively  (\citealt{margolis78,stecker79,michalak90}). 
The $pp$ cooling rate is almost independent of the proton energy and is given by (\citealt{kelner2006})
\begin{equation}\label{10}
t^{-1}_{pp}=n_i c \sigma_{pp}k_{pp},
\end{equation}
where, $n_i$ is the coronal number density of protons which can be calculated by eq.~\ref{nc}, and  $k_{pp}$ is the total inelasticity of the process of value $\sim 0.5$. The corresponding cross section for inelastic $pp$ interactions $\sigma_{pp}$ can be approximately by (\citealt{kelner2009})
\begin{equation}\label{11}
\sigma_{pp}(E_p)=\left(34.3+1.88Q + 0.25 Q^2\right)\left[1-\left(\frac{E_{th}}{E_p}\right)^4\right]^2 \rm mb,
\end{equation}
 where $\rm mb$  stands for milli-barn, $Q=\ln \left(\frac{E_p}{1TeV}\right)$, and the proton threshold kinetic energy for neutral pion $(\pi^0)$ production is $E_{th}=2m_\pi c^2(1+\frac{m_\pi}{4m_p})\approx280$ MeV, where $m_\pi c^2=134.97$ MeV is the rest energy of  $\pi^0$ (\citealt{vilaaharonian2009}). This particle decays in two photons with a probability of 98.8\%.

\begin{table}
\centering
\begin{minipage}{83mm}
\caption{\label{ta1} Three sets of model parameters for LLAGNs.}
\begin{tabular*}{\textwidth}{@{}llrrrrlrlr@{}}
\hline
 &Parameters &Model 1&Model 2&Model 3\\
\hline
$m$ &BH mass ($M_{\odot}$)&$10^7$&$10^8$&$10^9$\\
$p$&Injection spectral index&1.9&1.7&2.2\\
\hline
\end{tabular*}
\end{minipage}
\end{table} 


\subsection{$p\gamma$ interactions}

The photomeson ($p\gamma$) production takes place for photon energies greater than $E_{th}\approx$ 145MeV. Pions are also obtained from the $p\gamma$ interaction near the threshold via the channels (\citealt{atoyandermer2003})
\begin{equation}\label{12}
p+\gamma\rightarrow p+\pi^0, 
\end{equation}
with $\pi^0\rightarrow \gamma+\gamma$ carrying $20\%$ of accelerated protons energy and
\begin{equation}\label{13}
p+\gamma\rightarrow p+\pi^++\pi^-,
\end{equation}
and the charged pions will also decay producing neutrinos as described in \S. 3.1.

The radiative cooling rate for this mechanism in an isotropic photon field with density $n_{ph}(E_{ph})$ can be calculated by (\citealt{Stecker1968}):
\begin{dmath}\label{14}
t^{-1}_{p\gamma}(E_p)=\frac{c}{2\gamma_p^2}\int_{\frac{E_{th}^{(\pi)}}{2\gamma_p}}^\infty dE_{ph}\frac{n_{ph}(E_{ph})}{E^2_{ph}}\times \int_{E_{th}^{(\pi)}}^{2E_{ph}\gamma_p}d\epsilon_r \sigma_{p\gamma}^{(\pi)}(\epsilon_r)K_{p\gamma}^{(\pi)}(\epsilon_r)\epsilon_r,
\end{dmath}
where 
in our model the appropriate photons come from the synchrotron radiation
\footnote{We find that  for  photomeson production, the  radiation from the accretion disk is irrelevant compared to the contribution from the coronal synchrotron emission above the disk (see KGV15 and KGS15  for a detailed derivation of the synchrotron rate and its radiation field density).\\}, $n_{ph}(E_{ph})= n_{synch}(\epsilon)$,
$\gamma_p=\frac{E_p}{m_ec^2}$,
 $\epsilon_r$ is the photon energy in the rest frame of the proton, and $K_{p\gamma}^{(\pi)}$ is the inelasticity of the interaction. \cite{atoyandermer2003} proposed a simplified approach to calculate the cross-section and the inelasticity which are given by
\begin{equation}\label{15}
$$
\sigma_{p\gamma}(\epsilon_r)\approx \left\{ \begin{array}{rl}
340\   \mu barn &\mbox{$300 {\rm MeV}\leq \epsilon_r\leq {\rm 500 MeV}$} \\
120\  \mu barn &\mbox{ $\epsilon_r>500 {\rm MeV},$}
\end{array} \right.
$$
\end{equation}
and
\begin{equation}\label{16}
$$
K_{p\gamma}(\epsilon_r)\approx \left\{ \begin{array}{rl}
0.2\    &\mbox{$300 {\rm MeV}\leq \epsilon_r\leq 500 {\rm MeV}$} \\
0.6\   &\mbox{ $\epsilon_r>500 {\rm MeV}.$}
\end{array} \right.
$$
\end{equation}

\section{Neutrino emission and diffuse intensity}

To calculate the neutrino emission from  the nuclear region of an  LLAGN,  we consider a  population of  protons accelerated by magnetic reconnection in the surrounds of the BH according to the model described  in Section 2. 

We assume for these accelerated particles  an isotropic power law spectrum   (in units of $\rm{erg^{-1} {\rm cm^{-3} s^{-1}}}$)  (see e.g. KGV15 and references there in):
\begin{equation} \label{19}
Q(E)=Q_0 E^{-p}exp{[-E/E_{max}]}
\end{equation}
where $p>0$ and $E_{max}$ is the cut-off energy.

The normalization constant $Q_0$ above  is calculated from the total power injected to accelerate the  protons according to the relation:
\begin{equation} \label{18}
L_p=\int_{V} d^3r \int_{E_{min}}^{E_{max}}dE\ E\ Q(E)
\end{equation}
where $V$ is the volume of the emission region around the magnetic reconnection zone and  $L_p$ corresponds to the magnetic reconnection power $W$ given by Eq.~\ref{2}.
To calculate $W$  we have adopted the following suitable set of parameters $\xi=0.7$, $R_X=6 R_S$, $L_X=10 R_S$, and $L=20 R_S$. 

\begin{table*}
\centering
\begin{minipage}{117mm}
\caption{\label{ta2}Physical conditions around the LLAGNs represented by models 1, 2 and 3,  obtained from Eqs. 1 to 4, using $r_x=6$, $l=20$, $l_X=10$ and $\xi = 0.7$.}
\begin{tabular*}{\textwidth}{@{}llrrrrlrlr@{}}
\hline
 &Parameters &Model 1&Model 2&Model 3\\
\hline
$B$ &Magnetic field (\rm G)&$2.8\times10^4$&$8874$&$2806$\\
$W$&Magnetic reconnection power (\rm erg/s)&$2.4\times10^{42}$&$2.4\times10^{43}$&$2.4\times10^{44}$\\
$\Delta R_X$&Width of the current sheet (\rm cm)&$7.2\times10^{12}$&$7.2\times10^{13}$&$7.2\times10^{14}$\\
$n_c$&Coronal particle number density ($\rm cm ^{-3}$)&$3.6\times10^{10}$&$3.6\times10^{9}$&$3.6\times10^{8}$\\
\hline
\end{tabular*}
\end{minipage}
\end{table*}

The maximum energy of the accelerated particles $E_{max}$ is derived  from the balance between the magnetic reconnection acceleration rate (Eq.~\ref{5}) and the radiative loss rates as given in Section 3.
 Figure \ref{fig2}   compares these rates  for  protons considering LLAGNs with three different  BH masses  $10^7$, $10^8$ and $10^9 M_\odot$.  We have also considered different  power-law indices ($p$) for the injected particle spectrum in each of these models (see Table 1) which are  compatible with the values derived from analytical and numerical studies of first-order Fermi acceleration by magnetic reconnection and also with values inferred from the observations (e.g., KGS15). The calculated values of $B$, $W$, $\Delta R_X$ and $n_c$ from Eqs.~\ref{1}-\ref{nc} for these three  representative source models are listed in Table~\ref{ta2}.  
 For simplicity, we  consider the derived proton luminosities (which are $\sim 1/4 W$) and the emission properties of these three models to  characterize  the whole range of LLAGNs in  the calculation of the HE neutrino flux below.  The adoption of this approach, rather than accounting for a whole range of BH mass sources allows  us to  avoid the introduction of further free parameters in the modelling.  
 
In Fig.~\ref{fig2}, for comparison we have also calculated the proton  acceleration rate  due to a shock formed in the surrounds of the reconnection region (Eq.~\ref{7}) for the same set of parameters as above. As in KGV15 and KGS15, we find that the maximum energy attained from  magnetic reconnection acceleration is higher than that from the shock .    
It should be also remarked that  protons with these calculated maximum energies have Larmor radii smaller than the thickness of the reconnection layer $\Delta R_X$ (eq.~\ref{3}), as required.



The  neutrinos that are produced from pion decay will escape from the emission region without any absorption and their spectrum is given by (\citealt{tavecchio14,kimura14}):
\begin{equation} \label{20}
E_\nu L_\nu(E_\nu)\simeq (0.5t^{-1}_{pp}+\frac{3}{8} t^{-1}_{p\gamma}) \frac{L_X}{c} E_p L_p,
\end{equation}
where $E_\nu$ is the neutrino energy and $E_p$ the proton energy. 
 Since Figure~\ref{fig2} demonstrates that the $p\gamma$ emission cools the protons faster than $pp$ collisions, the dominant hadronic process in our model is the $p\gamma$ emission. Therefore,  this mechanism will prevail in the production  of the neutrinos and the first term of eq.~\ref{20} can be neglected. 
In $p\gamma$ interactions, $E_\nu$ is related with the parent proton energy through the equation $E_\nu = 0.05E_p$ (\citealt{spurio15}), because the average energy of the pion is $\sim 0.2$ of the parent proton energy and in the decay of the $\pi^+$ chain four leptons are produced (including one electron neutrino as remarked), each of which  has roughly 1/4 of the pion energy. 
It has been also demonstrated in \cite{spurio15} that the ratio of the neutrino luminosity to the photon luminosity from $p\gamma$ interactions is $\sim 1/3$.

 
In consistency with the statement above, the maximum energy of the produced  neutrinos  can be calculated from $E_{\nu,max}=0.05E_{p,max}$ (\citealt{becker08,hazlen07}), which according to our model is  $\sim 3\times10^{16} \rm{eV}$ for a source with a black hole mass $M_{BH}=10^7 M_{\odot}$, $\sim 5\times10^{15} \rm{eV}$  for a source with $M_{BH}=10^8 M_{\odot}$, and  $\sim 2\times10^{15} \rm{eV}$ for a source with $M_{BH}=10^9 M_{\odot}$.  

The total  diffuse neutrino intensity from the  extragalactic sources we are considering here, i.e., LLAGNs may have contributions from different redshifts. Neglecting evolutionary effects in the core region of these sources, we can estimate the total intensity as (\citealt{murase14}) 

\begin{dmath} \label{21}
\Phi_\nu=\frac{c}{4\pi H_0}\int_{0}^{z_{max}}dz \frac{1}{\sqrt{(1+z)^3\Omega_m+\Omega_\Lambda}}\times \int_{L_{min}}^{L_{max}} dL_{\gamma}\ \rho_{\gamma}(L_{\gamma},z)\frac{L_{\nu}(E_\nu)}{E_\nu},
\end{dmath}
where $L_\gamma$ is the $\gamma$-ray luminosity, and $\rho_{\gamma}(L_{\gamma},z)$ is the $\gamma$-ray luminosity function (GLF)
of the core of the sources, defined as the number density of sources per unit comoving volume, per unit logarithmic luminosity between the redshifts $z=0$ to $z=z_{max}$, being  the latter the maximum  observed redshift  for radiogalaxies, $z_{max}\simeq 5.2$  (\citealt{klamer05}). GLF is integrated  from $L_{min}$ to $L_{max}$ which are  obtained from \textit{Fermi}-LAT observations and are given by $10^{41}$ and $10^{44}$ erg/s, respectively (\citealt{dimauro14}).
The values for the cosmological parameters are assumed as: $H_0=70\ \rm{km \ s^{-1} Mpc^{-1}}$, $\Omega_M=0.3$ and $\Omega_\Lambda=0.7$. 
 
\begin{figure}
 \centering
 \includegraphics[width=3.5in]{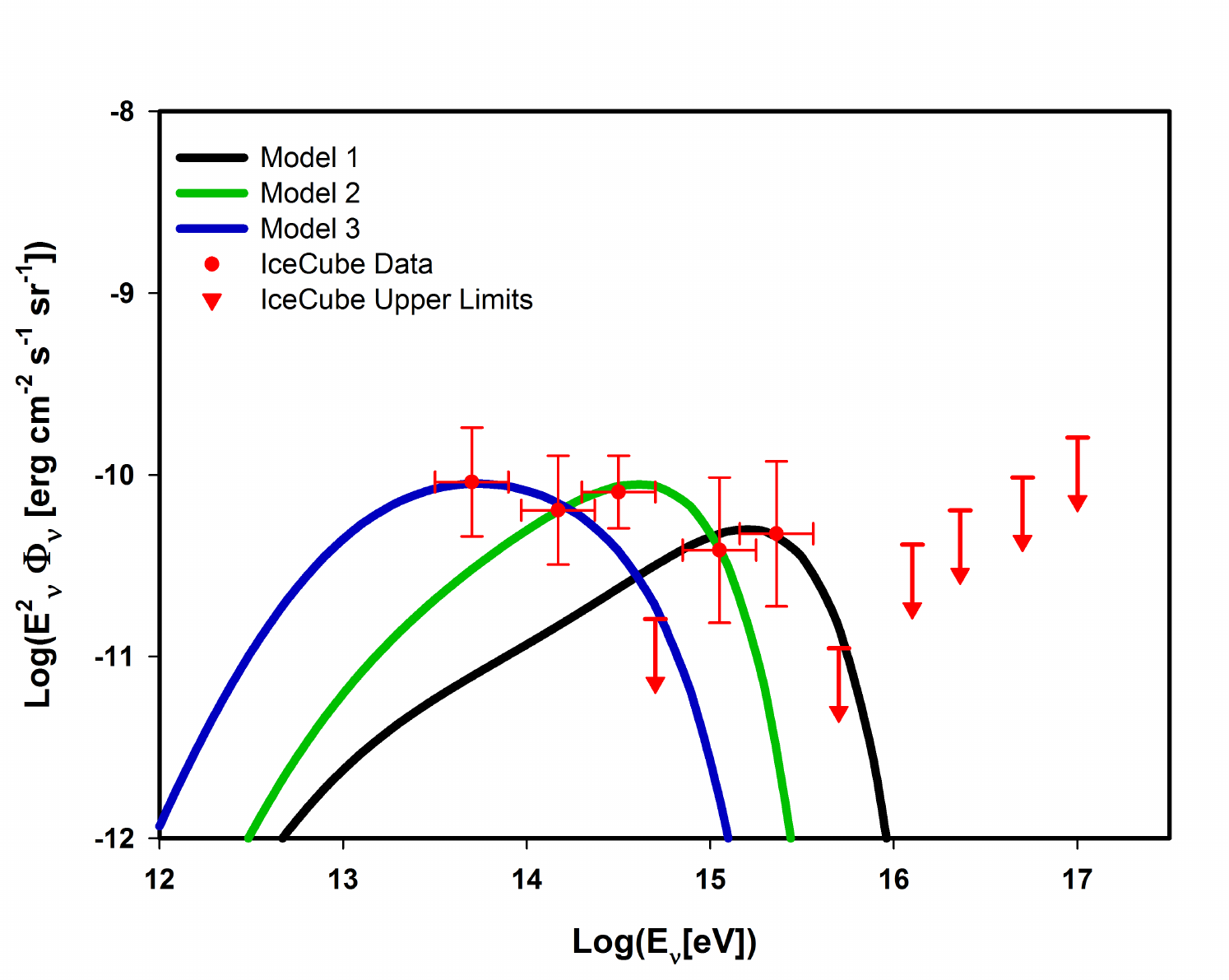}
 \caption{Calculated diffuse intensity of neutrinos from the cores of LLAGNs considering our magnetic reconnection acceleration model to produce the protons and  gamma-ray photons for three different BH masses. The data are taken from IceCube measurements (\citealt{aartsen14}).}
 \label{fig5}
\end{figure}

 We evaluate the GLF, $\rho_{\gamma}(L_{\gamma},z)$ as in \cite{dimauro14}, from the estimated radio luminosity function (RLF) which for non-blazars  is given by

 \begin{dmath} \label{22}
\rho_{\gamma}(L_{\gamma},z)=\rho_{r,tot}(L_{r,tot}^{5\rm{GHz}}(L_{r,core}^{5\rm{GHz}}(L_\gamma)),z)\times \frac{d\log{L_{r,core}^{5\rm{GHz}}}}{d \log{L_{\gamma}}}\ \frac{d\log{L_{r,tot}^{5\rm{GHz}}}}{d\log{L_{r,core}^{5\rm{GHz}}}}.
\end{dmath}
 ${d\log{L_{r,core}^{5\rm{GHz}}}}/{d \log{L_{\gamma}}}$ and ${d\log{L_{r,tot}^{5\rm{GHz}}}}/{d\log{L_{r,core}^{5\rm{GHz}}}}$ can be calculated by (\citealt{dimauro14})
\begin{equation} \label{23}
\log{L_\gamma}=2.00\pm 0.98+(1.008\pm 0.025)\log{(L_{r,core}^{5\rm{GHz}})}, 
\end{equation}
and
\begin{equation} \label{24}
\log{L_{r,core}^{5\rm{GHz}}}=4.2\pm 2.1+(0.77\pm 0.08)\log{(L_{r,tot}^{5\rm{GHz}})}, 
\end{equation}
 where $L_{r,tot}^{5\rm{GHz}}$ and $L_{r,core}^{5\rm{GHz}}$ are the radio total and core luminosities, respectively. 
 The total RLF, $\rho_{r,tot}(L_{r,tot}^{5\rm{GHz}}(L_{r,core}^{5\rm{GHz}}(L_\gamma)),z)$,  is found  from  interpolation of the observed data for radio-galaxies provided by \cite{yuanwang12}:
\begin{dmath} \label{25}
\rho_{r,tot}(L_{r,core}^{5\rm{GHz}}(L_{\gamma}),z)=(-1.1526\pm0.0411)\log{L_{r,core}^{5\rm{GHz}}}+(0.5947\pm0.1224)z+23.2943\pm1.0558\ \rm{Mpc^{-3}}( \log {L_{r,core}^{5\rm{GHz}}})^{-1}.
\end{dmath}

The resulting neutrino flux is shown in Figure~\ref{fig5}. It was calculated using eq. \ref{21} above, considering the maximum neutrino energies obtained for  sources with the three different BH masses (as in Figure 2).

Sources with $M_{BH}=10^7 M_{\odot}$ result a spectrum  that matches better with  the observed most energetic part of the  neutrino flux by the IceCube, at $\sim 3 \times 10^{15}$ to $10^{16}$ $\rm{eV}$, 
while  sources with BH masses of the order of $10^8 M_{\odot}$ produce a  spectrum that nearly fits  the observed  neutrinos flux in the range of $10^{14} - 10^{15} \rm{eV}$, and  sources with mass  $\sim10^9 M_{\odot}$ the narrow energy band  $5\times 10^{13}\rm{eV}-10^{14}\rm{eV}$ as well as the upper limit at $5\times10^{14}\rm{eV}$.

 \section{Discussion and Conclusions}
 
In this work we have explored a model to describe the observed flux  of extragalactic very high energy (VHE) neutrinos  by the IceCube (\citealt{aartsen14}) in the framework of low luminosity AGNs (LLAGNs), or more specifically, of radio-galaxies. The recent detection of  gamma-ray emission in the TeV range in these sources  makes them also potential candidates of  VHE neutrino emission via the decay  of charged pions which can be produced by the interaction of accelerated relativistic protons  with ambient lower energy photons and protons. 

We have examined here a fast    magnetic reconnection mechanism in the surrounds of the central BH occurring between the lines lifting from the accretion disk into the corona and those of the BH magnetosphere to accelerate particles to relativistic energies through a first-order Fermi process in the reconnection layer (GL05, \citealt{kowal2012}). Recently, it has been demonstrated that this  model   successfully reproduces the observed gamma-ray luminosity of hundreds of LLAGNs (\citealt{beteluis2014} and SGK15) and also shapes the  SEDs  of several radio-galaxies, particularly reproducing their TeV gamma-ray  energies mainly via photomeson  (p$\gamma$) interactions (KGS15).  

Applying the same acceleration model as above (see Section 2), considering three different BH masses,  we have shown that also the observed VHE neutrino Icecube flux (\citealt{aartsen14}) can be obtained from the decay of charged pions  produced in  photomeson interactions involving the accelerated protons and Synchrotron photons in the core region of these sources  (Figs. 2 and 3).

Specifically, in Fig.~\ref{fig2}, we compared  the magnetic reconnection acceleration rate (derived from the numerical simulations of \cite{kowal2012} and calculated for the source parameters)   with the relevant hadronic cooling processes and obtained the maximum energy for the accelerated protons mainly constrained by the  $p\gamma$  interactions. 
In Fig.~\ref{fig2}, we also compared the magnetic reconnection with the shock acceleration rate in the surrounds of the BH for the same parametric space and demonstrated the higher efficiency of the first process in this region. According to our results in Fig.~\ref{fig2}, protons are able to accelerate up to energies of the order of $\sim10^{17}\rm{eV}$ and therefore, are suitable to produce 0.1-1 PeV neutrinos.

 Fig.~\ref{fig5} indicates that the observed neutrino flux 
in the  few PeV range can be matched by  sources with $M_{BH}\sim 10^7 M_\odot$  (Model 1), while the flux  in the energy range of 0.1PeV$<E_\nu <$1PeV can be matched by sources  with $M_{BH}\sim 10^8 M_\odot$ (model 2),  and that in the range  $\leq 0.1\rm{PeV}$ can be fitted by sources with   $M_{BH}\sim 10^9 \rm{eV}$ (model 3). 

We note that, although the calculated neutrino  flux  was obtained from the integration of the contributions of LLAGNs over the redshifts between z=0 and 5.2 (eq. \ref{21}) considering, for simplicity, sources with only three characteristic values of BH masses, one may naturally  expect that a continuous integration considering the sources with all  possible BH masses within the range $10^7-10^9M_\odot$ should provide a similar fitting to the observed data.  We also note that our model is unable to explain the  IceCube upper limits at the $\sim 10\rm{PeV}$ range (also depicted in Fig.~\ref{fig5}), which are probably due to other astrophysical compact source population. 

Furthermore, we expect that with the 10-fold increased sensitivity at TeV energies, and the
larger field of view and improved angular resolution of the  forthcoming gamma-ray observatory CTA (\citealt{actis11,acharya13}), the  list of LLAGNs with confirmed detection of gamma-ray emission  at TeV energies (which currently has only four sources: Cen A, Per A, M87 an IC310),  will increase substantially, allowing for  a more precise evaluation of the contribution of individual sources  for the IceCube neutrino flux.

As remarked in \S. 1, other models have been proposed in the literature to explain the IceCube neutrino flux which cannot be discarded or confirmed, considering the current poorness of the data available. 

 \cite{tavecchio14} and \cite{tavecchio14b}, for instance,  have proposed  that the lower power blazar class of BL Lac objects could be promising candidates to produce the observed neutrino flux. In their two-zone jet model, the neutrinos are produced by photomeson interactions involving photons emitted in the slower, outer layer that envelopes the faster inner jet component. A limitation of this  model is that the high-energy cut-off of the accelerated protons, as well as their injected power are free parameters, unlike in our model where both quantities are directly obtained from the magnetic reconnection acceleration mechanism.   Besides,   since the BL Lacs are a subclass of the blazars, another  difficulty with this model  is that it is not clear  whether the remaining more powerful blazars, which are also TeV gamma-ray emitters,  can or cannot produce neutrinos. According to the recent studies of \cite{dermer14} and \cite{murase14}, which employed a single zone jet model, the  powerful blazars would not be suitable candidates to explain the IceCube data. These analyses and the relatively large number of free parameters employed  in the evaluation of the neutrino flux leave the question on  whether or not  blazars do contribute to the IceCube data opened.

Another model to explain the observed neutrino flux has been proposed by \cite{kalashev14} who studied   photo-pion production  on the anisotropic photon field of a Shakura-Sunyaev  accretion disk in the vicinity of the BH in AGNs. But this model does not provide an acceleration mechanism either and therefore, the proton high energy cut-off  is also a free parameter.

Recently, radio galaxies have been also discussed as possible sources of the observed  HE neutrinos  by Becker et al. (2014). 
They  demonstrated that FR I radio galaxies  would be more probable sources of this emission than FR II radio galaxies. In this work, as we considered the global diffuse contribution from LLAGNs  spread over a range of $z$ values, we cannot distinguish the relative contributions from both classes.

Finally, another recent study  (\citealt{kimura14}) speculates that the protons responsible for the neutrino emission could be accelerated stochastically by the turbulence induced in a RIAF accretion disk in the core region of LLAGNs. This acceleration process should be essentially a second-order Fermi process and therefore, less efficient than a first-order Fermi process. Nevertheless,  their  analytically estimated acceleration rate $t^{-1}_{acc}\propto E^{-0.35}$ seems to be too large when compared to that predicted for first-order Fermi processes, as for instance in the present study where the acceleration rate has been extracted directly from 3D MHD simulations with test particles ($t^{-1}_{acc}\propto E^{-0.4}$; \citealt{kowal2012}, KGV15), or in shock acceleration (for which analytic predictions give $t^{-1}_{acc}\propto E^{-1}$ (\citealt{spruit88})). Furthermore,  since the candidates to produce PeV neutrinos in this case are radio-galaxies, which are the observed $\gamma$-ray emitters, it seems that the employment of the gamma-ray luminosity function (GLF; \citealt{dimauro14})  to calculate the diffuse neutrino intensity as we did here seems to be more appropriate than  the   employment of the luminosity function in X-rays, as these authors considered.

In summary, in spite of its simplicity,  the  numerically tested acceleration model  
applied to the core region of LLAGNs here presented indicates that LLAGNs are very promising candidates to explain the   IceCube VHE neutrinos.

\section*{Acknowledgements}
This work has been partially supported by grants of the Brazilian agencies FAPESP (2013-10559-5, and 2011/53275-4), CNPq (306598/2009-4) and CAPES. We are also indebted to the
 referee for his/her useful comments.

\bsp

\label{lastpage}

\end{document}